\documentstyle[11pt,twoside,newpasp,epsf]{article}
\renewcommand{\thepage}{\arabic{page}}
\begin{document}

\title{An Empirical Ultraviolet Iron Spectrum Template Applicable to Active Galaxies}
\author{M.\ Vestergaard}
\affil{Dept. of Astronomy, The Ohio State University, Columbus, OH 43210}

\author{B.\ Wilkes}
\affil{Smithsonian Astrophysical Observatory, Cambridge, MA 02138}

\begin{abstract}
Iron emission is often a severe contaminant in optical-ultra\-violet 
spectra of active galaxies. Its presence complicates emission line studies.  
A viable solution, already successfully applied at optical wavelengths, is 
to use an empirical iron emission template. We have generated Fe\,{\sc ii} 
and Fe\,{\sc iii} templates for ultraviolet active galaxy spectra based on 
{\it HST} archival 1100\,--\,3100\,\AA{} spectra of I\,Zw\,1. 
Their application allows fitting and subtraction of the iron emission in
active galaxy spectra.  This work has shown that in particular 
C\,{\sc iii}]\,$\lambda$1909 can be heavily contaminated by other line
emission, including iron transitions.  Details of the data processing,
generation, and use of the templates, are given by Vestergaard \& Wilkes (2001).

\end{abstract}

\section{Motivation}

This work was motivated by our desire to study the intrinsically emitted ultraviolet 
(UV) broad line profiles of active galactic nuclei (AGNs) to get more accurate and
reliable information on e.g., the chemical abundances and dynamical and ionization
structure in their broad line region.

In addition to the commonly studied strong, permitted lines of 
Ly\,$\alpha \,\lambda$\,1216, Si\,{\sc iv}+O\,{\sc iv}]\,$\lambda$\,1400,
C\,{\sc iv}\,$\lambda$\,1549, C\,{\sc iii}]\,$\lambda$\,1909, and 
Mg\,{\sc ii}\,$\lambda$\,2800, AGN restframe-UV spectra also contain weaker 
lines such as various transitions of carbon, silicon, oxygen, nitrogen, helium, 
and iron.  The importance of these lines lies not only in their presence and 
strengths but also in their effective ``contamination'' of the stronger lines.  
Iron is by far the strongest of these ``contaminants'' of the UV (and optical) 
emission lines. Due to the large number of electron levels in iron atoms, 
thousands of emission-line transitions of several ionization states are 
distributed throughout the UV and optical spectral regions (e.g., Kurucz 1997).  
The weak lines of 
iron and other metals blend together, especially when the AGN intrinsic line 
width is large. This heavy blending of weak, broad lines, dominated by iron 
transitions, forms a {\em pseudo-continuum} above the intrinsically emitted
continuum level (Wills, Netzer \& Wills 1985) even in high signal-to-noise AGN
spectra. The presence of such a pseudo-continuum severely complicates the study 
of the wings of the prominent resonance lines, the weak line features themselves, 
and the intrinsically emitted continuum emission. 
It is important to make reliable measurements of these features in order 
to study issues such as continuum\,--\,line relationships, dynamical and 
ionization structure, and chemical abundances in the AGN broad line region.

Our preliminary analysis indicates that line measurement uncertainties of order 
$\sim$5\,--\,25\% are expected in the prominent lines if the iron emission is not 
corrected for. Some weak emission lines may, however, be so severely blended with
iron emission that the uncertainty approaches 100\%.

\section{Why an Empirical Template?}

A viable way to correct for the contaminating iron emission is to generate an
empirical iron emission template which can be used to `model' the iron emission in
individual AGN spectra. An alternative is to theoretically model the iron emission. 
However, to date such modeling has only been moderately successful mainly due to 
inadequate atomic data and limited understanding of the rather complicated iron 
emission mechanism. Much improved atomic data have become available in recent years 
(e.g., Hummer et al. 1993; Seaton et al. 1994; Kurucz 1997) 
allowing much more advanced theoretical models to be developed. 
This work is currently in progress (e.g. E.\ Verner 2001, private communication;
Verner et al. 1999).
An empirical iron template is thus extremely useful until adequate theoretical 
models and an understanding of the iron spectrum emitted by AGNs
are available.  A template is not model dependent but comprises the iron emission 
transitions and line ratios as emitted by an active galaxy.  It permits the 
fitting and subtraction of the iron emission in individual AGN spectra, as the 
template can be broadened and scaled to match the iron emission in each AGN
spectrum.  In addition, this allows us to create a database of (fitted) 
iron spectra which can help constrain theoretical models of AGN iron emission, 
thereby helping us to better understand this emission.
An optical template, based on data of I\,Zw\,1, has already been successfully 
applied to the bright quasar survey (Boroson \& Green 1992).

\begin{figure}
\plotfiddle{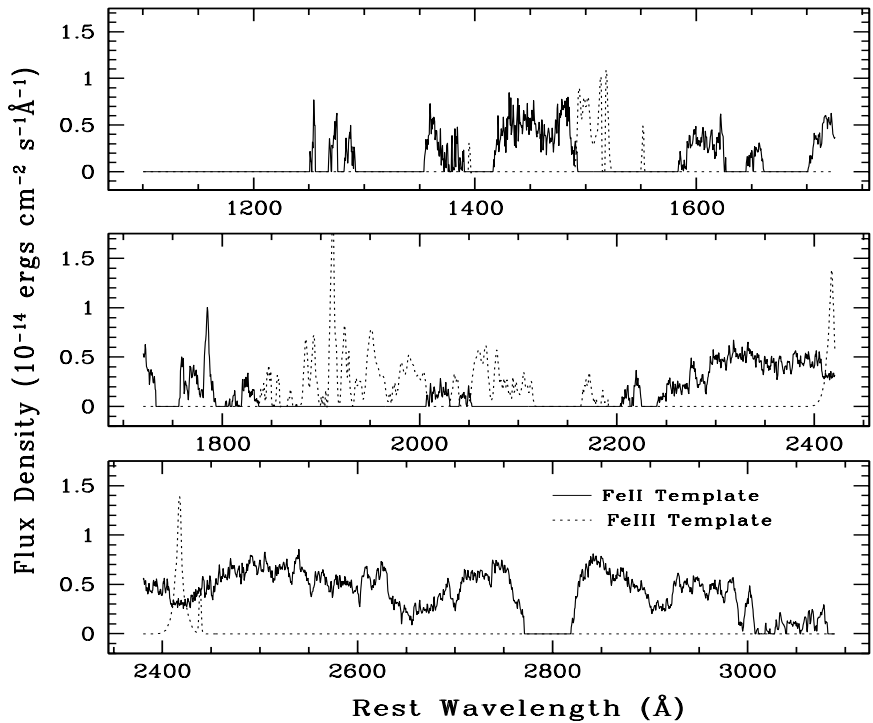}{8.59cm}{0}{65}{54}{-205}{-105}
\caption{The UV iron template. The Fe\,{\sc ii} emission is shown by the solid line,
while Fe\,{\sc iii} is shown by the dotted line. \label{fig-1}}
\end{figure}

\section{The Template}

We present a UV Fe\,{\sc ii} and Fe\,{\sc iii} empirical template ranging
from 1250\,--\,3090\,\AA\ based on high quality {\it HST} archival spectra of 
the nearby AGN, I\,Zw\,1.
The data were originally obtained, presented, and analyzed by Laor et al.
(1997), who also identify most of the emission and absorption features in the 
spectra.  The UV spectrum of I\,Zw\,1 was chosen for several reasons. 
The spectrum is rich and strong in the iron emission, the prominent resonance 
lines are relatively narrow (full width at half maximum $\sim$\,900\,km\,s$^{-1}$) 
and an optical iron emission template based on this object already exists, as 
mentioned.

A strong and rich iron spectrum permits the detection of weak iron features and 
the identification of as many iron features as are presumably typically emitted
by AGNs. The relatively narrow lines allow us to deblend the line emission in 
the process of generating the iron spectrum.
In order to isolate the iron emission, the other 
emission and absorption lines were modeled (composite Gaussian profiles 
were used to reproduce the line profile shapes) and subtracted.

The narrow I\,Zw\,1 line width reveals that a number of line-emission 
transitions may dominate the C\,{\sc iii}]\,$\lambda$\,1909 line blend; 
the C\,{\sc iii}] contribution is small for I\,Zw\,1. 
Hence, caution must be exercised when studying C\,{\sc iii}] as 
direct measurements of this line can be significantly in error for broader 
lined active galaxies where the emission lines blend together and are 
difficult to discern.  This affects general line studies, broad-line region density 
determinations, and abundance studies. This heavy blending of C\,{\sc iii}] 
emphasizes the motivation for this work.

The Fe\,{\sc ii} and Fe\,{\sc iii} emission templates are shown in 
Figure~\ref{fig-1}. Small residuals of the subtracted line fits are set 
to zero to avoid introducing unnecessary uncertainties into the AGN spectra 
to which the templates are applied. In the region around Ly\,$\alpha$ the 
iron emission is either too weak to be reliably isolated or non-existent, 
except in the red wing of Ly\,$\alpha$.  Blueward of Ly\,$\alpha$ the Lyman 
forest heavily complicates a reliable isolation of the iron emission. 
The strength of the Fe\,{\sc iii} $\lambda$\,1540\AA\ component is very 
uncertain as it is heavily blended with C\,{\sc iv} emission, and blueshifted 
C\,{\sc iv} emission is possibly present.  Vestergaard \& Wilkes (2001) 
present details of the template generation, its application, and address 
the limitations of the template fitting method.

\section{Future Work}

The Fe\,{\sc ii} and Fe\,{\sc iii} templates are used to fit the UV iron emission 
in large samples of quasar spectra (Forster et al.\ 2001; Vestergaard et al.\ 2001,
in preparation) 
and can be used to fit any AGN UV spectrum with intrinsic 
line width $\geq$900\,km\,s$^{-1}$.

The fitting and subtraction of the contaminating iron emission in UV spectra of
AGN will improve studies of their broad emission lines and 
will help improve photoionization modeling of the broad line region.
The database of fitted iron emission spectra from individual active galaxies 
can then be used to attempt to understand this iron emission in terms of 
physical models.

\acknowledgments

We are grateful to Dima Verner for discussions and direction to
updated iron line lists and to Ari Laor for useful discussions.
MV is very pleased to thank the Smithsonian Astrophysical Observatory, 
where most of this work was done, for their hospitality and gratefully 
acknowledges financial support from the Columbus Fellowship, the Danish
Natural Sciences Research Council (SNF-9300575), the Danish Research Academy
(DFA-S930201), and a Research Assistantship at Smithsonian Astrophysical
Observatory (NAGW-4266, NAGW-3134, NAG5-4089).  BJW gratefully acknowledges
financial support from NASA contract NAS8-39073 (Chandra X-ray Center).

\end{document}